\documentclass[twocolumn,showpacs,preprintnumbers,amsmath,amssymb]{revtex4}

\usepackage{graphicx}
\usepackage{dcolumn}
\usepackage{bm}

\begin{document}

\title{Comment on ``Fluctuation-induced first-order transition $p$-wave superconductors" \\ by Qi Li, D. Belitz, and J. Toner [Phys. Rev. B {\bf 79}, 054514 (2009)]}

\author{Dimo I. Uzunov}

\altaffiliation[]{Electronic address: d.i.uzunov@gmail.com}
\affiliation{Collective Phenomena Laboratory,
 Institute of Solid State Physics, Bulgarian Academy of Sciences, BG--1784, Sofia, Bulgaria.
}

\date{2nd July 2006}
\begin{abstract} In this Comment, we show that the paper by Qi Li, D. Belitz and J. Toner, published in Phys. Rev. B {\bf 79}, 054514 (2009), contains an incomplete mean-field analysis of a simple model of Ginzburg-Landau type. The latter contains a stable non-unitary phase, which has not been found in this study and is missing in the outlined picture of possible stable phases. In this Comment, the mean field analysis has been corrected, the errors have been explained in details and relevant topics have been discussed. Shortcomings in the mean-field-like and renormalization group studies in the mentioned paper have also been revealed.
\end{abstract}

\pacs{64.60.ae,64.60.De, 74.20.De}

\keywords{unconventional superconductivity, critical fluctuations, phase transition, renormalization group}

\maketitle

A paper by Qi Li {\it et al.}~\cite{QiLi:2009} is intended to present new results about global and local gauge effects on the phase transitions in unconventional ({\em p}-wave) superconductors. The study in Ref.~\cite{QiLi:2009} is based on the same methods as those applied in Refs.~\cite{Blagoeva:1990, Millev:1990, Busiello:1991, Volovik:1985} (see also Refs.~\cite{Uzunov:2010, Halperin:1974, Busiello:1986}): (i) mean-field (MF) approximation~\cite{Uzunov:2010}, (ii) MF-like approximation ~\cite{Uzunov:2010, Halperin:1974}, and (iii) a renormalization group (RG) investigation within the one-loop approximation combined with $\epsilon$-expansion to first order in $\epsilon = (4-d)$; $d$ is the spatial dimensionality~\cite{Uzunov:2010}. As in previous studies~\cite{Blagoeva:1990, Millev:1990, Busiello:1991}, the main tasks in Ref.~\cite{QiLi:2009} are MF derivation of the possible phases and RG study of the so-called ``fluctuation-induced weakly-first-order phase transition"\cite{Halperin:1974} in $p$-wave superconductors.

 In this Comment we present irrefutable arguments for essential shortcomings of Ref.~\cite{QiLi:2009}. Substantial errors are corrected. Related topics are elucidated. The reader must know the correct results and the history of the problem.

 The authors of Ref.~\cite{QiLi:2009} conducted a MF analysis of the free energy density
\begin{equation}
\label{Eq1}
 f(\mbox{\boldmath$\psi$})  = t|\mbox{\boldmath$\psi$}|^2 + u|\mbox{\boldmath$\psi$}|^4 + v|\mbox{\boldmath$\psi$}\times\mbox{\boldmath$\psi$}^{\ast}|^2,
\end{equation}
\noindent
(cf.~Eq. (3.1) in Ref.~\cite{QiLi:2009}), where $t$, $u$, and $v$ are vertex (Landau) parameters, and $\mbox{\boldmath$\psi$} = (\psi_1,\psi_2,\psi_3)$ is a complex vector field (the order parameter of $p$-wave superconductors). They used an appropriate representation
\begin{equation}
\label{Eq2}
 \mbox{\boldmath$\psi$} =  \psi_0 (\hat{n}\cos\phi + i\hat{m}\sin\phi)
\end{equation}
\noindent
by the modulus $\psi_0 \equiv |\mbox{\boldmath$\psi$}|$, the auxiliary (``phase") angle $\phi$, and the unit vectors $\hat{n}$ and $\hat{m}$ ($\hat{n}.\hat{m} = \cos\theta$). The unit vectors $\hat{n}$ and $\hat{m}$ point out the directions of the real and imaginary vector parts of the field $\mbox{\boldmath$\psi$} = (\mbox{\boldmath$\psi$}^{\prime}+ i\mbox{\boldmath$\psi$}^{\prime\prime}$): $\hat{n}\parallel \mbox{\boldmath$\psi$}^{\prime}$, and $\hat{m} \parallel \mbox{\boldmath$\psi$}^{\prime\prime}$.

The authors of Ref.~\cite{QiLi:2009} correctly claim that $f(\mbox{\boldmath$\psi$})$ is minimized by $\hat{n}=\hat{m}$ and $\psi_0^2=  -t/2u$, provided $v > 0$, $u > 0$, and $t < 0$. The minimum is $f(t,u) = -t^2/4u$ and, as mentioned in Ref.~\cite{QiLi:2009}, describes the unitary phase -- the phase, characterized by the property $\mbox{\boldmath$\psi$}\times \mbox{\boldmath$\psi$}^{\ast} = 0$. According to these authors, in the case of $v>0$, this is the only available stable order ($\mbox{\boldmath$\psi$} \neq 0$).

This outline of stable phases in superconductors with $v>0$ and described by model~(\ref{Eq1}) is incorrect, as we shall show in the remainder of this Comment. First, the description of the unitary phase is incomplete. Second, there exists another phase -- a non-unitary phase, which is stable when $v > 0$.

In fact, the careful analysis of the equations of state $\partial f/\partial\psi_0 = \partial f/\partial\theta =\partial f/\partial\phi = 0$ inevitably leads to two topologically different domains~\cite{remark:2012} of the unitary phase ($\mbox{\boldmath$\psi$}\times\mbox{\boldmath$\psi$}^{\ast} = 0$), as given by the actual minimizing condition $\hat{n}= \pm\hat{m}$, i.e., $\theta = \pi l$, where $l = 0, \pm 1, \dots $ (in Ref.~\cite{QiLi:2009}, this condition is presented in a wrong way: $\hat{n}= \hat{m}$, i.e., $\theta = 2\pi l$). The two domains of the same unitary phase are distinguished by the parallel and antiparallel mutual orientations of $\hat{n}$ and $\hat{m}$. This remark is important for studies of more realistic models, where the crystal field anisotropy is taken into account (for previous studies of this type, see Refs.~\cite{Blagoeva:1990, Millev:1990, Busiello:1991, Volovik:1985}).

The stability matrix $\hat{S} =\{S_{ij}\} = (\partial^2f/\partial \mu_i\partial \mu_j)$, where $\mathbf{\mu} = {\mu_i} = (\psi_0,\theta,\phi)$, describes the stability of the possible phases (equilibria) with respect to variations of the order parameters ($\psi_0,\theta,\phi$) around their equilibrium values. For the unitary phase ($\mbox{\boldmath$\psi$}\times\mbox{\boldmath$\psi$}^{\ast} = 0$), this matrix has the form
\begin{equation}
\label{Eq3}
 \hat{S} = \left[ \begin{array}{ccc}
 -4t, & 0,& 0\\
 0,& \left(vt^2/2u^2\right)\sin^{2}2\phi, & 0\\
 0,&0,&0
 \end{array}\right].
\end{equation}
\noindent
Using Eq.~(\ref{Eq3}) and the existence conditions ($t < 0$, $u > 0$), one obtains the stability conditions $u > 0$, $v> 0$, $t>0$, as indicated in Ref.~\cite{QiLi:2009}.

Besides, there are additional circumstances, which should be emphasized. It is readily seen from Eq.~(\ref{Eq3}) that the unitary phase is marginally stable with respect to $\theta$-fluctuations for angles $\phi = \pi k/2$ ($k = 0,\pm 1, \dots)$, whereas the same phase has a marginal stability towards $\phi$-fluctuations for any $\theta$ (in both cases, $t < 0$, $u >0$, and $v > 0$). As shown in previous works, see~\cite{Blagoeva:1990, Volovik:1985} and references therein, these marginal stabilities towards the angles $\theta$ and $\phi$ reflect the global gauge invariance of the model~(\ref{Eq1}) and the phenomenon of global symmetry breaking in the six-dimensional space ($\Re\psi_1, \Im\psi_1,\dots, \Re\psi_3,\Im\psi_3 $) of the order parameter vector $\mbox{\boldmath$\psi$}$. Thus the zeros of the matrix elements $S_{22}$ and $S_{33}$, produced by some values of $\theta$ and $\phi$, exhibit the fundamental global $SO(3)$ symmetry of model (\ref{Eq1}) rather than an uncertainty in the stability degree of the unitary phase (the same is valid for the non-unitary phases, discussed below; see also~\cite{Blagoeva:1990, Volovik:1985}).

Ref.~\cite{QiLi:2009} reports only for a particular type of non-unitary phase ($\mbox{\boldmath$\psi$}\times\mbox{\boldmath$\psi$}^{\ast} \neq 0$), where $\hat{n} \perp \hat{m}$, i.e., $\theta = \pi(l + 1/2)$; existence and stability conditions: $(u+v) > 0$, $v < 0$, and $t < 0$. The result for $\theta$ follows as a direct solution of the equations of state. Obviously, this phase has the property $\mbox{\boldmath$\psi$}^{\prime} \perp \mbox{\boldmath$\psi$}^{\prime\prime}$. In Sec. I--III as well as anywhere in the paper~\cite{QiLi:2009}, there are no references to previous papers with MF results for the phases of model (\ref{Eq1}), although this model is often used (see Refs.~\cite{Blagoeva:1990, Millev:1990, Busiello:1991, Volovik:1985, Uzunov:2010}, and references therein). Moreover, the authors do not make any stipulation for some purposive incompleteness of their consideration; neither in Sec. III.A nor elsewhere. Thus the reader remains with the wrong conclusion that the non-unitary phase of type $\mbox{\boldmath$\psi$}\perp \mbox{\boldmath$\psi$}^{\ast}$ is the only non-unitary phase ($\mbox{\boldmath$\psi$}\times \mbox{\boldmath$\psi$}^{\ast} \neq 0$) described by the model (\ref{Eq1}).

In fact, the careful MF analysis~\cite{Blagoeva:1990} shows that the model (\ref{Eq1}) contains a stable non-unitary phase in more.  This {\it second} non-unitary phase (2ndNUP) is given by
\begin{equation}
\label{Eq4}
\psi_0^2= -\frac{t}{2u}, \;\;\;\; \theta \neq \pi l, \;\;\;\; \phi = \frac{\pi}{2}k,
\end{equation}
\noindent ($k = 0, \pm 1,\dots $), the free energy minimum $f = -t^2/4u$ -- the same as for the unitary phase, and the stability matrix
\begin{equation}
\label{Eq5}
 \hat{S} = \left[ \begin{array}{ccc}
 -4t, & 0,& 0\\
 0,& 0, & 0\\
 0, &0, &\left(2vt^2/u^2\right)\sin^2\theta
 \end{array}\right].
\end{equation}
\noindent

The 2ndNUP has the same modulus $\psi_0$ and energy $f$ as the unitary phase. The difference with the unitary phase is in the restriction $\phi = (\pi/2)k$ as well as in the equilibrium values of angle $\theta$. The Eq.~(\ref{Eq5}) shows that the second non-unitary phase is stable for $t < 0$,  $u>0$, and $v > 0$ -- stability conditions, which are identical to those for the unitary phase. In the limit $\theta \rightarrow \pi l$ the structure and the stability properties of 2ndNUP coincide with those of the unitary phase. Having in mind the same free energies of these phases, one may conclude that the unitary phase and the second non-unitary phase are identical for $\theta \rightarrow \pi l$. These results show that the $u-v$ phase diagram outlined in Fig.~1 of Ref.~\cite{QiLi:2009} is wrong and should be corrected along our instructions.

The authors of Ref.~\cite{QiLi:2009} have missed to find the 2ndNUP, contained in model (\ref{Eq1}). On the other hand, they have missed to take advantage from paper~\cite{Blagoeva:1990} (and references therein), where this phase is described in MF (in slightly different notations; see Sec.~II and especially, Sec. II.D.4 in Ref.~\cite{Blagoeva:1990}). In Refs.~\cite{Blagoeva:1990, Volovik:1985}, the possible $u-v$ phase diagrams of the free energy (\ref{Eq1}) have been outlined in details as a particular case of a quite more general description and more realistic variants of such  diagrams. 

In Sec. III.B of Ref.~\cite{QiLi:2009}, the Eq.~(3.5) for the MF-like renormalization of the free energy $f(\mbox{\boldmath$\psi$})$ contains an essential error. In this equation, the parameter $w$ should be related to the charge $q$ by $w \sim q^3 $ rather than by the wrong relationship $w \sim q$, given on the line below the equation. In order to persuade himself in this, one should have a look on the original article~\cite{Halperin:1974} or some other sources, for example, Ref.~\cite{Uzunov:2010}.

Besides, within the MF-like approximation, in which the spatial fluctuations of the field $\mbox{\boldmath$\psi$}$ are neglected but the fluctuations of the magnetic induction in the entire Ginzburg-Landau fluctuation Hamiltonian are preserved ~\cite{Uzunov:2010, Halperin:1974}, the parameters $t$ and $u$ acquire $q$-renormalizations, too. The latter are important for the description of essential properties of the weakly-first-order phase transition~\cite{Uzunov:2010}. Note, that in accord with Eq.~(2e) in Ref.~\cite{Busiello:1991}, the parameter $v$ does not possess $q$-renormalizations within the model (\ref{Eq1}), neither in the one-loop RG treatment nor within the MF-like approximation.

In Sec.~III.C--D of Ref.~\cite{QiLi:2009}, for the aim of study of spatial fluctuations of $\mbox{\boldmath$\psi$}$, the approximation of uniform field [$\mbox{\boldmath$\psi$}(\mathbf{x}) \approx \mbox{\boldmath$\psi$}$] is abandoned and the $\mathbf{x}$-variations of $\mbox{\boldmath$\psi$}$ are taken into account. Within this extended scheme, the critical fluctuations of the field $\mbox{\boldmath$\psi$}(\mathbf{x})$ can be studied with the help of RG. Here we briefly discuss the RG analysis of the ``action" (2.3), performed in Ref.~\cite{QiLi:2009} (Sec.III.C) within the one-loop approximation and first-order $\epsilon$-expansion. Note, that the mentioned level of accuracy in the application of RG is enough to reveal the main features of the fluctuation effects on the properties of the phase transition from the normal metal state to the $p$-wave superconducting state at the phase transition  temperature, where the mean magnetic induction is equal to zero, and a particular type of fluctuation-induced weakly-first-order phase transition may appear~\cite{Uzunov:2010, Halperin:1974}.

First, we note, that Eq.~(2.3) does not present an action, as wrongly noted in Ref.~\cite{QiLi:2009}. Rather, Eq.~(2.3) presents a ``generalized free energy (functional)" (alias, a ``fluctuation Hamiltonian" rather than an ``action"). The same is valid for the quantity $S_A$, given by Eq.~(4a) and describing the time-independent spatial fluctuations of the magnetic induction for the particular case of uniform $\mbox{\boldmath$\psi$}$.

Second, the RG analysis within the $\epsilon$-expansion performed in Ref.~\cite{QiLi:2009} is an entire repetition of the RG study in Ref.~\cite{Millev:1990}. Apart of some slight differences in the notations, the RG Eqs.~(3.7a)--(3.7c) are the same as the first three equations (3) in Ref.~\cite{Millev:1990}. Note, that owing to a simple misprint in Ref.~\cite{Millev:1990}, the number factor 12 in the last term of the equation for $du/dl$ should be 96. This misprint is without any consequence for the results derived in Ref.~\cite{Millev:1990}. A second note is needed to point out that in Ref.~\cite{Millev:1990} we use known RG equations for the parameters $c$, $q$, and $\mu$ (Refs.~\cite{Halperin:1974, Busiello:1986}). These equations are exactly the same as Eqs.~(3.7d)--(3.7f) in Ref.~\cite{QiLi:2009}. They do not contain the parameter $v$ and are known from the original paper~\cite{Halperin:1974} as well as from numerous further publications; see, e.g., Refs.~\cite{Uzunov:2010, Busiello:1986}.

Thus our RG equations, presented in Ref.~\cite{Millev:1990} and the RG equations, re-derived in Ref.~\cite{QiLi:2009} are the same. Then it is not strange that the authors of Ref.~\cite{QiLi:2009} have rediscovered our RG results about the possible fixed points and their stability properties, as well as about the possible types of critical behavior and first-order phase transition.

For example, using Eqs.~(3.11a) and (3.11b), the authors of Ref.~\cite{QiLi:2009} rediscover the couple of fixed points of type $u_{\pm} \neq 0$, $\tilde{u}_{\pm} \neq 0$, $v_{\pm} =0$ (notations of Ref.~\cite{Millev:1990}, where $\tilde{u}$ corresponds to $v$ in Ref.~\cite{QiLi:2009}; the fixed point coordinates $v_{\pm}$ correspond to the crystal anisotropy parameter considered there but neglected in Ref.~\cite{QiLi:2009}). Moreover, the condition for the existence of this couple of fixed points, $ m > m_b = n_b/2 = 211$, derived for the first time in Ref.~\cite{Millev:1990} has also been re-derived in Ref.~\cite{QiLi:2009} in a slightly different notation: $2m=n > n_c^{P} \approx 420.9$ (in a minor difference with the result in Ref.~\cite{Millev:1990}). Let us remind, that within the present problem, the symmetry index $n$ is an integer equal to the total number of component of the complex vector $\mbox{\boldmath$\psi$}$, while $m=n/2$ is equal to the number of components of the real vectors $\mbox{\boldmath$\psi$}^{\prime}$ and $\mbox{\boldmath$\psi$}^{\prime\prime}$. Following Ref.~\cite{QiLi:2009}, in our discussion of MF results for the phases, we have used $n=2m = 6$. But in the RG studies large integer $n$ are considered, too.

The main conclusions of Refs.~\cite{QiLi:2009} and ~\cite{Millev:1990} are the same: The critical behavior appears for $n > n_b$. For $n < n_b$, the phase transition from normal metal state to $p$-wave superconductivity is a fluctuation-induced weakly-first-order phase transition. Under suitable physical conditions~\cite{Uzunov:2010}, this weakly-first-order phase transition may appear in the known superconductors, where $n$ does not exceed $6$.

The difference in the $\epsilon$-analysis performed in these two papers is in the wider scope of Ref.~\cite{Millev:1990}, where the crystal anisotropy has been taken into account, and the effect of this anisotropy on the stability of the critical behavior has been investigated in details. The crystal anisotropy enhances the effect of first-order transition. For example, in the present of crystal anisotropy, the critical behavior appear for $m = 2n > 5494$ (see Ref.\cite{Millev:1990}). Certainly, in real metals and metallic compounds, the crystal anisotropy effect on the $p$-wave superconductivity cannot be neglected.

Moreover, the RG analysis in Ref.~\cite{QiLi:2009} is a limited case of the RG study presented in Ref.~\cite{Busiello:1991}. In Ref.~\cite{Busiello:1991}, the simultaneous effect of crystal anisotropy and quenched impurities has been obtained by a generalization of the simple fluctuation model (2.1)--(2.3), given in Ref.~\cite{QiLi:2009}. The simple check of article contents undoubtedly demonstrates that the RG equations in Ref.~\cite{QiLi:2009} are a limited case of those in Ref.~\cite{Busiello:1991} and describe a simplified picture, where both crystal anisotropy and quenched impurity effects are ignored. Hence, the RG analysis and the respective RG results in Ref.~\cite{QiLi:2009} exactly follow from the RG study in Ref.~\cite{Busiello:1991} when the parameters, describing the crystal anisotropy and quenched disorder are neglected.

Finally, we summarize our main findings. In Ref.~\cite{QiLi:2009}, the MF analysis is incomplete and a stable non-unitary phase is missed. In result, the domain of stability of this phase should be indicated in the $u-v$ phase diagram, shown in Fig.~(1). Excepting this error, the MF analysis in Ref.~\cite{QiLi:2009} repeats previous studies and re-discover known results (see Ref.~\cite{Blagoeva:1990} and references therein). The parameter $w$ in Sec.III.B should be corrected, as shown in this Comment. The RG study, performed in Ref.~\cite{QiLi:2009} within the $\epsilon$ expansion, totally repeats essential parts of previous RG studies (Refs.~\cite{Millev:1990, Busiello:1991}) and, hence, re-derives known results.

\end{document}